\documentclass[11pt]{article} 
\usepackage{rldmsubmit,palatino}
\usepackage{graphicx}

\title{Sequence Modeling for N-Agent Ad Hoc Teamwork}

\author{
Caroline ~Wang\thanks{Equal contribution.} \\
Department of Computer Science\\
The University of Texas at Austin\\
Austin, TX 78712 \\
\texttt{caroline.l.wang@utexas.edu} \\
\And
Di Yang ~Shi\footnotemark[1] \\
Department of Computer Science\\
The University of Texas at Austin\\
Austin, TX 78712 \\
\texttt{dyshi@utexas.edu} \\
\AND
Elad Liebman \\
Amazon\thanks{Work done outside of Amazon} \\
\texttt{liebelad@amazon.com} \\
\And
Ishan Durugkar \\
Sony AI \\
\texttt{ishan.durugkar@sony.com} \\
\And
Arrasy Rahman\\
Department of Computer Science\\
The University of Texas at Austin\\
Austin, TX 78712 \\
\texttt{arrasy@cs.utexas.edu} \\
\And
Peter Stone \\
Department of Computer Science \\
The University of Texas at Austin and Sony AI \\
\texttt{pstone@cs.utexas.edu} \\
}

\usepackage[numbers]{natbib}
\usepackage{amssymb}
\usepackage{amsmath}
\usepackage{amsthm}
\usepackage{enumitem}
\usepackage{bm}
\usepackage{bbm}
\usepackage{wrapfig}
\usepackage{graphicx}
\usepackage{xcolor} 
\usepackage{wrapfig}

\newcommand{\citetemp}[1]{}

\begin{document}

\maketitle

\begin{abstract}
N-agent ad hoc teamwork (NAHT) is a newly introduced challenge in multi-agent reinforcement learning, where controlled subteams of varying sizes must dynamically collaborate with varying numbers and types of unknown teammates without pre-coordination. 
The existing learning algorithm (POAM) considers only independent learning for its flexibility in dealing with a changing number of agents. 
However, independent learning fails to fully capture the inter-agent dynamics essential for effective collaboration. 
Based on our observation that transformers deal effectively with sequences with varying lengths and have been shown to be highly effective for a variety of machine learning problems, this work introduces a centralized, transformer-based method for N-agent ad hoc teamwork. 
Our proposed approach incorporates historical observations and actions of all controlled agents, enabling optimal responses to diverse and unseen teammates in partially observable environments. Empirical evaluation on a StarCraft II task demonstrates that MAT-NAHT outperforms POAM, achieving superior sample efficiency and generalization, without auxiliary agent-modeling objectives.
\end{abstract}

\keywords{
multi-agent reinforcement learning, ad hoc teamwork, transformers, agent modeling
}

\acknowledgements{This work has taken place in the Learning Agents Research Group (LARG) at UT Austin.  LARG research is supported in part by NSF (FAIN-2313998, FAIN-2019844, NRT-2125858), ONR (N00014-18-2243), ARO (W911NF-23-2-0004, W911NF-17-2-0181), DARPA (Cooperative Agreement
HR00112520004 on Ad Hoc Teamwork) Lockheed Martin, and UT Austin's Good Systems grand challenge.  Peter Stone serves as the Executive Director of Sony AI America and receives financial compensation for
this work.  The terms of this arrangement have been reviewed and approved by the University of Texas at Austin in accordance with its policy on objectivity in research.
}  
\startmain 

\section{Introduction \& Background}
\label{sec:intro-bg}

Cooperative multi-agent reinforcement learning (CMARL) is a paradigm for learning coordinated team policies on fully cooperative tasks, assuming full control over all agents. In contrast, ad hoc teamwork (AHT) considers the challenge of training a single agent to collaborate with previously unknown teammates on fully cooperative tasks.

The recently introduced N-agent ad hoc teamwork (NAHT) problem generalizes both paradigms by considering the scenario where a set of \textit{controlled} agents must collaborate with dynamically varying numbers and types of \textit{uncontrolled} teammates without the ability to pre-coordinate. The primary challenges of the NAHT problem are (1) dealing with a changing number of (un)controlled agents, and (2) generalizing to unseen uncontrolled teammate behavior \citep{wang2024naht}.

As a first solution method for NAHT, \citet{wang2024naht} proposed the Policy Optimization with Agent Modeling (POAM) algorithm. POAM deals with a potentially varying number of controlled agents by employing an independent policy gradient approach, where each agent learns an independent policy while performing teammate modeling of all other agents in the team. Each agent's policy conditions on the learned teammate representations, thus enabling the policy to approximate the optimal behavior for collaborating with each type of uncontrolled team.
While independent learning allows dealing with a changing number of controlled agents, the independent policy structure does not enable learning all possible joint policies, limiting the richness of team behaviors. This work instead considers a centralized learning approach that may still deal with a changing number of controlled agents by leveraging a transformer architecture.

Transformers have achieved state-of-the-art performance in sequence modeling tasks across language and vision domains \citep{brown2020gpt3, dosovitskiy2021vit, vaswani2017attn}. Their ability to handle variable-length inputs and capture complex dependencies has inspired applications in reinforcement learning (RL) as well. For instance, Decision Transformers frame offline RL as a sequence modeling task in which a transformer predicts actions based on historical observations and rewards \citep{chen2022dt}.

Transformer-based architectures have also proven highly successful in CMARL.
\citet{kuba2022happo} proposed a multi-agent advantage decomposition theorem, which decomposes the joint (team) advantage by defining an arbitrary ordering over agents. Next, the joint advantage can be decomposed into a sum of per-agent advantages, where each agent's advantage depends on the actions selected by prior agents in the ordering.
Motivated by the sequential nature of the multi-agent advantage decomposition theorem, \citet{wen2022multi} framed CMARL as a sequential \textit{decision-making} problem, and proposed a transformer-based method for CMARL, called the Multi Agent Transformer (MAT). 
The MAT is a fully centralized approach to CMARL, that has demonstrated state-of-the-art performance in various MARL benchmarks. 
The key idea of MAT is the sequential decision-making paradigm, in which some ordering over teammates is defined, and each agent in the ordering makes decisions based on the actions of the previous agents in the ordering.
The sequential decision-making architecture allows efficient multi-agent policy optimization in a fully centralized setting, that scales linearly in the number of agents rather than exponentially.
By learning a fully centralized value function, the MAT also avoids the need to make limiting value factorization assumptions \citep{rashid18qmix, son_qtran_2019, sunehag18vdn}.

The MAT was originally designed for CMARL tasks, where the team size is fixed and the algorithm has full control over all agents. Further, it does not consider history, limiting its applicability for partially observable problems such as NAHT. However, based on our observation that the sequential decision-making architecture is suitable for controlling a varying number of agents, this work considers a centralized, multi-agent transformer architecture for the NAHT setting,  which we call MAT-NAHT. 
The MAT-NAHT architecture performs centralized decision-making for a varying number of controlled agents. The controlled subteam is able to respond optimally to diverse and unknown teammates, using historical observations and actions of controlled agents. We find that MAT-NAHT demonstrates superior performance over POAM in StarCraft II tasks, without explicit agent-modeling auxiliary objectives.


\section{Preliminaries}
\label{sec:prelim}
The N-agent ad hoc teamwork (NAHT) problem is formalized in the framework of Decentralized Partially Observable Markov Decision Processes (Dec-POMDPs). 
A Dec-POMDP is described by the tuple $\langle M, S, \{O^i\}_{i=1}^M, A, T, r, \gamma, T \rangle$. 
Here, $M$ denotes the number of agents, $S$ the global state space, $A = \prod_{i=1}^{M}A^i$ the joint action space, and $O^i$ the local observation space for agent $i$. 
Let the notation $\Delta(\bullet)$ denote a probability distribution over the space $\bullet$.
The transition function $T: S \times A \rightarrow \Delta(S)$ governs the dynamics, and the reward function $r: S \times A \rightarrow \mathbb{R}$ provides a shared reward signal. 
In this work, we examine a centralized control scheme where each agent selects actions according to the stochastic policy $\pi^i: H^i \times A^1 \times \cdots A^{i-1} \rightarrow \Delta(A^i)$, where $H^i$ is the history of observations and actions available to agent $i$, and the policy conditions on the actions selected by agents $1$ to $i-1$ in some fixed, arbitrary ordering.
The decision-making problem occurs under the horizon $T$.

The NAHT problem considers a controlled set of agents $C$,  and an uncontrolled set $U$. The learning algorithm generates policies for agents in $C$, while uncontrolled agents in $U$ follow fixed, potentially unknown policies. Each episode begins with a team sampling process $X(U, C)$, consisting of two steps: first, the number of controlled agents $N < M$ is sampled, and second, a set of $N$ agents is drawn from the controlled set $C$, while $M-N$ agents are drawn from the uncontrolled set $U$. The NAHT learning problem considers a parameterized set of controlled agents, $C(\theta)$, and aims to maximize the expected return under varying teammates and team compositions: $\max_{\theta} \mathbb{E}_{\{\pi^{i}\}_{i=1}^M \sim X(U, C(\theta))} \left[ \sum_{t=0}^{T} \gamma^t r_t \right].$

\paragraph{Training and Evaluation in Practice}
Similar to the training and evaluation procedure specified by \citet{wang2024naht}, we assume access to a training and test set of teammates, $X_{\text{train}}$ and $X_{\text{test}}$, which are drawn from the same distribution over teammate types. NAHT algorithms learn from interaction with teammates within the training set $X_{\text{train}}$, while evaluations are conducted with teammates in the test set, $X_{\text{test}}$.

\section{Multi-Agent Transformers for N-Agent Ad Hoc Teamwork}
\label{sec:method}

\begin{figure}
    \centering
    \includegraphics[width=1\linewidth]{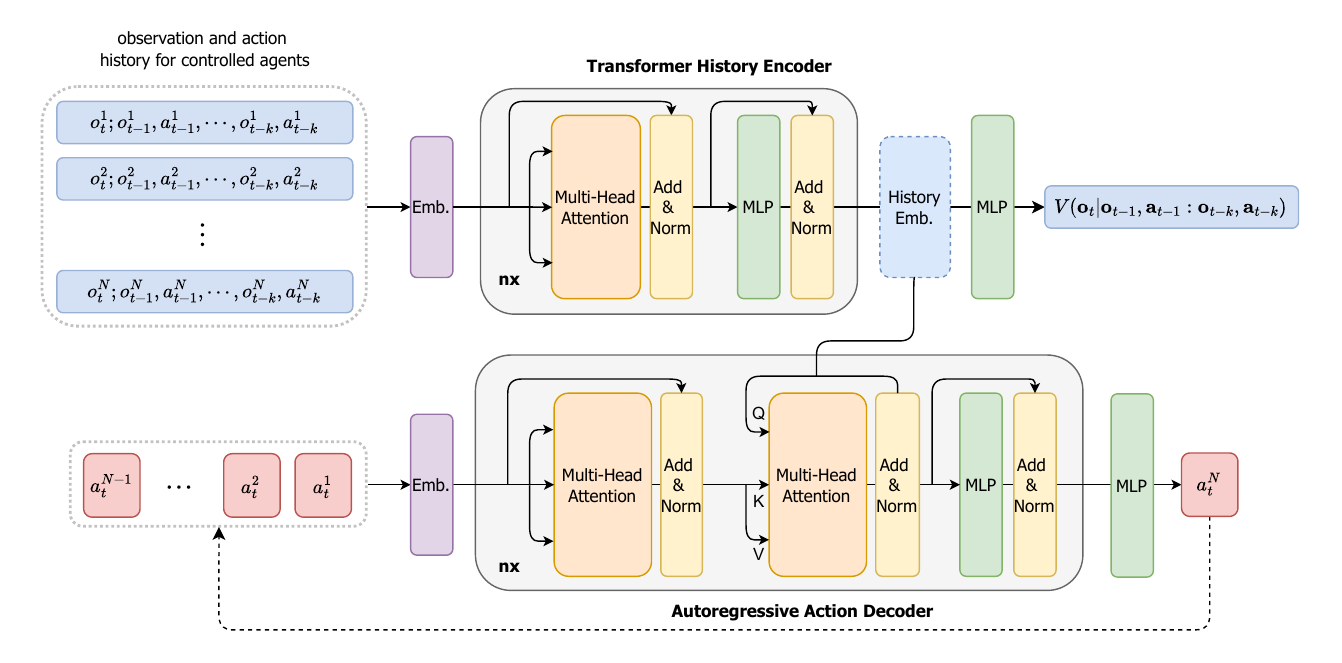}
    \caption{Architecture of the multi-agent transformer for NAHT. Only observations and actions corresponding to the controlled agents (agents $1, \cdots, N$) are provided to the encoder. Similarly, the autoregressive action decoder decodes actions for the controlled agents only.}
    \label{fig:method-architecture}
\end{figure}

The NAHT problem requires (1) dealing with varying numbers of controlled agents and (2) adapting to diverse types of unknown teammates without prior coordination. 
To address the first requirement, we use a transformer-based architecture, for its ability to handle variable input sizes and complex dependencies effectively.
To address the second, decision making is conditioned on the history of observations and actions, to allow implicit teammate behavior modeling.

Our proposed method, MAT-NAHT, follows the MAT architecture with modifications tailored to the NAHT problem. 
An encoder-decoder transformer \citep{vaswani2017attn} acts as both the value and policy network for the controlled team, and is trained via the PPO algorithm \citep{Schulman2017ProximalPO}. 
The the encoder-decoder transformer receives information from, and specifies actions for, the controlled agents only. During training, the restriction of information to controlled agents is implemented by masking out data corresponding to the uncontrolled agents.  
We assume that the observations of the controlled agents contain sufficient information about the uncontrolled teammates to enable controlled agents to recognize and respond to uncontrolled teammates. 
MAT-NAHT's architecture is illustrated in Figure~\ref{fig:method-architecture} and described in further detail below. 

\paragraph{Encoder.} 
The encoder processes the current observation and the history of observations and actions for all controlled agents, for the last $k$ timesteps of history. 
Using an approach similar to the Decision Transformer \citep{chen2022dt}, the sequence of agent observation histories is augmented by learned positional encodings, embedded, and then processed using a multi-head self-attention mechanism to generate representations of the current and historical observation and actions for all controlled agents.

The learned history representations then play two roles. First, they are passed to the decoder, where they act as the query when performing autoregressive action decoding. Second, the representations are further processed into history-conditioned, joint value predictions, $V(\mathbf{o}_t |\mathbf{o}_{t-1}, \mathbf{a}_{t-1}: \mathbf{o}_{t-k}, \mathbf{a}_{t-k} )$.

\paragraph{Decoder.} 
Following \citet{wen2022multi}, the autoregressive decoder generates actions for each controlled agent sequentially, conditioned on the history representations and previous actions of controlled agents. The action generation follows a fixed, but arbitrary order. 

\begin{wrapfigure}{R}{0.3\textwidth} 
    \vspace{-50pt}
    \includegraphics[width=0.3\textwidth]{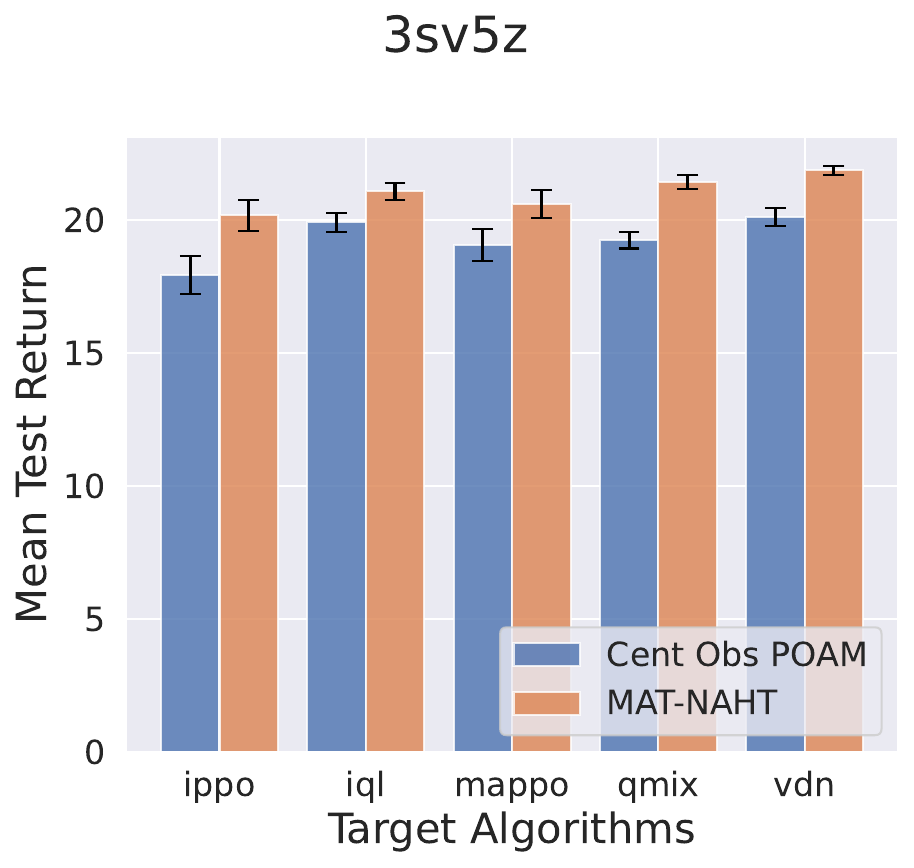}
    \caption{Mean test return and 95\% CI on 3sv5z for MAT-NAHT versus Cent Obs POAM when paired with $X_{test}$ teammates, computed over five trials. MAT-NAHT has improved generalization compared to Cent Obs POAM, across all five test teammate types.}
    \vspace{-40pt}
    \label{fig:3sv5z_ood}
\end{wrapfigure}

\section{Experimental Evaluation}
\label{sec:exp}

This section presents an initial empirical evaluation of the MAT-NAHT method. We investigate the following question: Does MAT-NAHT improve over POAM, given multiple types of unseen and uncontrolled teammates? 


\subsection{Experimental Setting}

We perform experiments on the StarCraft Multi-Agent Challenge (SMAC) benchmark \citep{samvelyan19smac}, which provides a number of cooperative multi-agent tasks in the setting of the popular real-time strategy game, StarCraft II. The tasks consist of allied teams,  which are controlled by a learned team policy, battling enemy teams, which are controlled by a fixed game AI.

We select the 3s vs 5z (\texttt{3sv5z}) task for our initial study, which has been considered a hard task in the literature \citep{wen2022multi, yu2021surprising}. This task 
consists of three allied stalker units versus five enemy zealot units.
To win the battle in the \texttt{3sv5z} task, the allied agents must learn a mechanic known as ``kiting"---where agents alternate movement and attack actions to maximize damage dealt to the enemy, while minimizing damage taken. 

\paragraph{Environment Details}
The observation space of the SMAC environment consists of the relative $x$ and $y$ positions of allied and enemy units, relative distances, shield health, and unit-type for all units visible within the ego agent's field of view. It also includes nearby terrain features corresponding to height and walkability, and the previous action of all visible units.
The action space is discrete and consists of the four cardinal movement directions and a no-op action. Dead agents can only take no-op action while live agents cannot. 
Finally, the reward function consists of a term specifying whether the battle was won or lost, and shaping terms corresponding to the damage dealt or received, and the number of dead allied and enemy units.

\begin{wrapfigure}{L}{0.35\textwidth} 
    \vspace{-10pt}
    \includegraphics[width=0.35\textwidth]{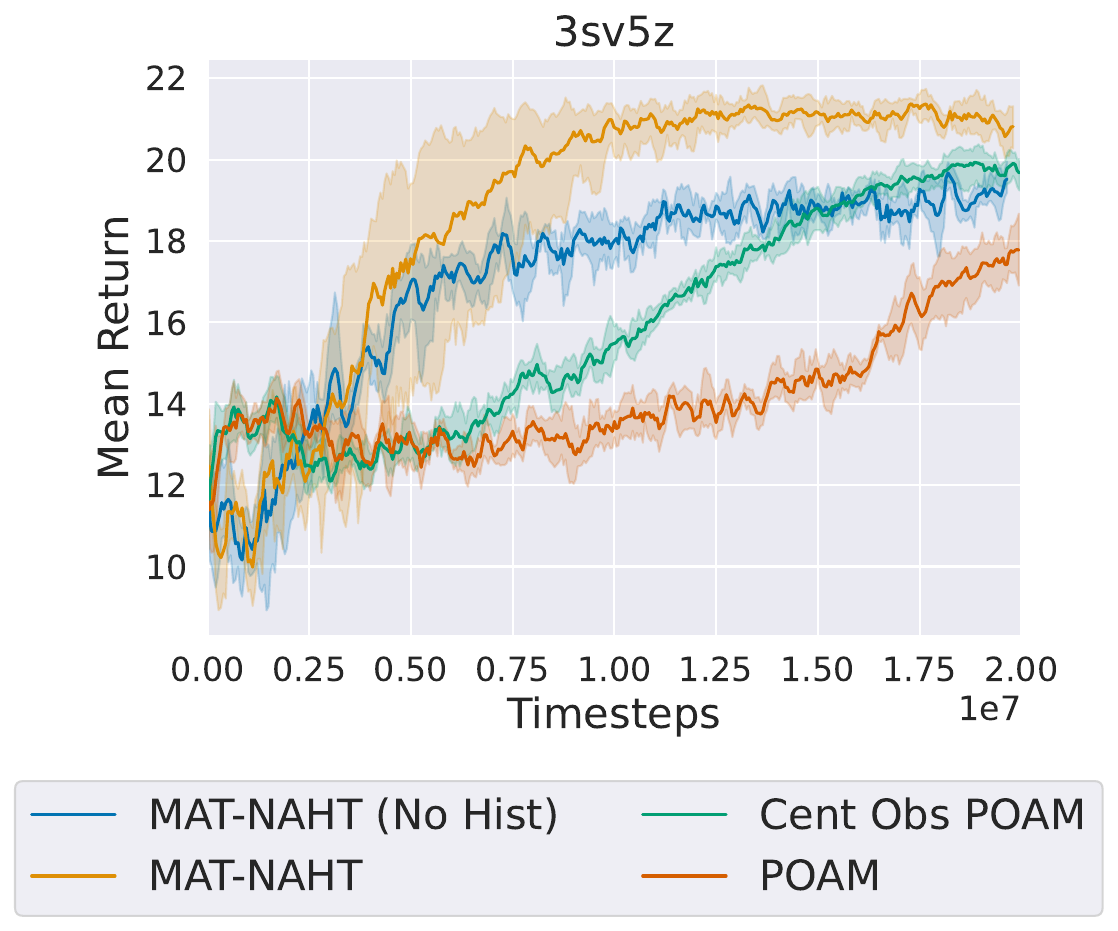}
    \caption{Sample efficiency curves with a 95\% CI on \texttt{3sv5z} task across five trials. This reflects the performance over $X_{train}$ teammates averaged over the uncontrolled teammate types.}
    \vspace{-40pt}
    \label{fig:3sv5z_sample_eff}
\end{wrapfigure}

\paragraph{Uncontrolled Teammates}
Following the procedure outlined in \citet{wang2024naht}, we train five types of uncontrolled teams using five MARL algorithms: VDN, IPPO, MAPPO, IQL, and QMIX \cite{sunehag18vdn, yu2021surprising, Tan1997MultiAgentRL, rashid18qmix}. For a particular episode with $M$ total agents, an uncontrolled team is first sampled uniformly at random from the set of five team types. Next,  the number of controlled agents is sampled uniformly, $N \in[1, M-1]$. In the episode, $N$ controlled agents must coordinate with $M-N$ uncontrolled teammates of the sampled type. 

\paragraph{Baseline}
The only prior algorithm for NAHT is POAM, which is a fully independent learning method. 
To enable a fairer comparison to MAT-NAHT, which employs a fully centralized critic, we introduce a centralized version of POAM (Cent Obs POAM) that provides the observations of all controlled teammates to the critic.  
Figure~\ref{fig:3sv5z_sample_eff} shows that Cent Obs POAM improves over POAM in terms of sample efficiency, leading to an approximate improvement of 10\% within the training budget. 


\subsection{Results}
Figure~\ref{fig:3sv5z_ood} shows the test returns of MAT-NAHT and Cent Obs POAM, when evaluated with the test set of uncontrolled teams, $X_{\text{test}}$. MAT-NAHT achieves higher test returns than Cent Obs POAM, demonstrating improved generalization. 
Moreover, Figure~\ref{fig:3sv5z_sample_eff} shows that at training time, MAT-NAHT attains significantly higher sample efficiency compared to all other evaluated methods---for instance, converging in less than half the timesteps of Cent Obs POAM. 

We also performed an ablation study removing the observation history from MAT-NAHT. Figure~\ref{fig:3sv5z_sample_eff} shows that MAT-NAHT without history achieves a lower returns than MAT-NAHT with history, within the training budget with history. 
We hypothesize that MAT-NAHT's performance drops without history because it is not possible for a policy to distinguish between multiple uncontrolled teammate types,  without access to history. 
Nevertheless, MAT-NAHT without history has similar learning  efficiency to Cent Obs POAM. 

\section{Discussion}
Transformer architectures has proven remarkably successful in a large variety of settings.
This abstract presents initial findings on applying a centralized, transformer-based method for decision-making in N-agent ad hoc teamwork. 
Initial empirical evaluations on a StarCraft II task show that the our method (MAT-NAHT) generalizes better than baselines. 
In future work, we plan to extend evaluation to further tasks, and analyze whether trained MAT-NAHT models are implicitly performing agent modeling. 
We also plan to consider alternative  architectures, such as spacio-temporal transformers or state-space models, to enable decision-making with longer historical context over multiple episodes.

\bibliographystyle{abbrvnat} 
\bibliography{refs}

\begin{thebibliography}{14}
\providecommand{\natexlab}[1]{#1}
\providecommand{\url}[1]{\texttt{#1}}
\expandafter\ifx\csname urlstyle\endcsname\relax
  \providecommand{\doi}[1]{doi: #1}\else
  \providecommand{\doi}{doi: \begingroup \urlstyle{rm}\Url}\fi

\bibitem[Brown et~al.(2020)Brown, Mann, Ryder, Subbiah, Kaplan, Dhariwal, Neelakantan, Shyam, Sastry, Askell, Agarwal, Herbert-Voss, Krueger, Henighan, Child, Ramesh, Ziegler, Wu, Winter, Hesse, Chen, Sigler, Litwin, Gray, Chess, Clark, Berner, McCandlish, Radford, Sutskever, and Amodei]{brown2020gpt3}
T.~Brown, B.~Mann, N.~Ryder, M.~Subbiah, J.~D. Kaplan, P.~Dhariwal, A.~Neelakantan, P.~Shyam, G.~Sastry, A.~Askell, S.~Agarwal, A.~Herbert-Voss, G.~Krueger, T.~Henighan, R.~Child, A.~Ramesh, D.~Ziegler, J.~Wu, C.~Winter, C.~Hesse, M.~Chen, E.~Sigler, M.~Litwin, S.~Gray, B.~Chess, J.~Clark, C.~Berner, S.~McCandlish, A.~Radford, I.~Sutskever, and D.~Amodei.
\newblock Language models are few-shot learners.
\newblock In H.~Larochelle, M.~Ranzato, R.~Hadsell, M.~Balcan, and H.~Lin, editors, \emph{Advances in Neural Information Processing Systems}, volume~33, pages 1877--1901. Curran Associates, Inc., 2020.

\bibitem[Chen et~al.(2021)Chen, Lu, Rajeswaran, Lee, Grover, Laskin, Abbeel, Srinivas, and Mordatch]{chen2022dt}
L.~Chen, K.~Lu, A.~Rajeswaran, K.~Lee, A.~Grover, M.~Laskin, P.~Abbeel, A.~Srinivas, and I.~Mordatch.
\newblock Decision transformer: reinforcement learning via sequence modeling.
\newblock In \emph{Advances in Neural Information Processing Systems (NeurIPS)}, 2021.

\bibitem[Dosovitskiy et~al.(2021)Dosovitskiy, Beyer, Kolesnikov, Weissenborn, Zhai, Unterthiner, Dehghani, Minderer, Heigold, Gelly, Uszkoreit, and Houlsby]{dosovitskiy2021vit}
A.~Dosovitskiy, L.~Beyer, A.~Kolesnikov, D.~Weissenborn, X.~Zhai, T.~Unterthiner, M.~Dehghani, M.~Minderer, G.~Heigold, S.~Gelly, J.~Uszkoreit, and N.~Houlsby.
\newblock An image is worth 16x16 words: Transformers for image recognition at scale.
\newblock In \emph{International Conference on Learning Representations}, 2021.
\newblock URL \url{https://openreview.net/forum?id=YicbFdNTTy}.

\bibitem[Kuba et~al.(2022)Kuba, Chen, Wen, Wen, Sun, Wang, and Yang]{kuba2022happo}
J.~G. Kuba, R.~Chen, M.~Wen, Y.~Wen, F.~Sun, J.~Wang, and Y.~Yang.
\newblock Trust region policy optimization in multi-agent reinforcement learning.
\newblock In \emph{International Conference on Learning Representations}, 2022.

\bibitem[Rashid et~al.(2018)Rashid, Samvelyan, Schroeder, Farquhar, Foerster, and Whiteson]{rashid18qmix}
T.~Rashid, M.~Samvelyan, C.~Schroeder, G.~Farquhar, J.~Foerster, and S.~Whiteson.
\newblock {QMIX}: Monotonic value function factorisation for deep multi-agent reinforcement learning.
\newblock In \emph{Proceedings of the 35th International Conference on Machine Learning}, volume~80 of \emph{Proceedings of Machine Learning Research}. PMLR, 2018.

\bibitem[Samvelyan et~al.(2019)Samvelyan, Rashid, de~Witt, Farquhar, Nardelli, Rudner, Hung, Torr, Foerster, and Whiteson]{samvelyan19smac}
M.~Samvelyan, T.~Rashid, C.~S. de~Witt, G.~Farquhar, N.~Nardelli, T.~G.~J. Rudner, C.-M. Hung, P.~H.~S. Torr, J.~Foerster, and S.~Whiteson.
\newblock {The} {StarCraft} {Multi}-{Agent} {Challenge}.
\newblock \emph{CoRR}, abs/1902.04043, 2019.

\bibitem[Schulman et~al.(2017)Schulman, Wolski, Dhariwal, Radford, and Klimov]{Schulman2017ProximalPO}
J.~Schulman, F.~Wolski, P.~Dhariwal, A.~Radford, and O.~Klimov.
\newblock Proximal policy optimization algorithms.
\newblock \emph{ArXiv}, abs/1707.06347, 2017.
\newblock URL \url{https://api.semanticscholar.org/CorpusID:28695052}.

\bibitem[Son et~al.(2019)Son, Kim, Kang, Hostallero, and Yi]{son_qtran_2019}
K.~Son, D.~Kim, W.~J. Kang, D.~E. Hostallero, and Y.~Yi.
\newblock {QTRAN}: {Learning} to {Factorize} with {Transformation} for {Cooperative} {Multi}-{Agent} {Reinforcement} {Learning}.
\newblock In \emph{Proceedings of the 36th {International} {Conference} on {Machine} {Learning}}, pages 5887--5896. PMLR, May 2019.
\newblock URL \url{https://proceedings.mlr.press/v97/son19a.html}.

\bibitem[Sunehag et~al.(2018)Sunehag, Lever, Gruslys, Czarnecki, Zambaldi, Jaderberg, Lanctot, Sonnerat, Leibo, Tuyls, and Graepel]{sunehag18vdn}
P.~Sunehag, G.~Lever, A.~Gruslys, W.~M. Czarnecki, V.~Zambaldi, M.~Jaderberg, M.~Lanctot, N.~Sonnerat, J.~Z. Leibo, K.~Tuyls, and T.~Graepel.
\newblock Value-decomposition networks for cooperative multi-agent learning based on team reward.
\newblock In \emph{Proceedings of the 17th International Conference on Autonomous Agents and Multi Agent Systems}, AAMAS '18, 2018.

\bibitem[Tan(1997)]{Tan1997MultiAgentRL}
M.~Tan.
\newblock Multi-agent reinforcement learning: Independent versus cooperative agents.
\newblock In \emph{International Conference on Machine Learning}, 1997.
\newblock URL \url{https://api.semanticscholar.org/CorpusID:267858156}.

\bibitem[Vaswani et~al.(2017)Vaswani, Shazeer, Parmar, Uszkoreit, Jones, Gomez, Kaiser, and Polosukhin]{vaswani2017attn}
A.~Vaswani, N.~Shazeer, N.~Parmar, J.~Uszkoreit, L.~Jones, A.~N. Gomez, L.~u. Kaiser, and I.~Polosukhin.
\newblock Attention is all you need.
\newblock In I.~Guyon, U.~V. Luxburg, S.~Bengio, H.~Wallach, R.~Fergus, S.~Vishwanathan, and R.~Garnett, editors, \emph{Advances in Neural Information Processing Systems}, volume~30. Curran Associates, Inc., 2017.

\bibitem[Wang et~al.(2024)Wang, Rahman, Durugkar, Liebman, and Stone]{wang2024naht}
C.~Wang, A.~Rahman, I.~Durugkar, E.~Liebman, and P.~Stone.
\newblock N-agent ad hoc teamwork.
\newblock In \emph{Advances in Neural Information Processing Systems (NeurIPS)}, 2024.

\bibitem[Wen et~al.(2022)Wen, Kuba, Lin, Zhang, Wen, Wang, and Yang]{wen2022multi}
M.~Wen, J.~Kuba, R.~Lin, W.~Zhang, Y.~Wen, J.~Wang, and Y.~Yang.
\newblock Multi-agent reinforcement learning is a sequence modeling problem.
\newblock In \emph{Advances in Neural Information Processing Systems}, volume~35, 2022.

\bibitem[Yu et~al.(2022)Yu, Velu, Vinitsky, Wang, Bayen, and Wu]{yu2021surprising}
C.~Yu, A.~Velu, E.~Vinitsky, Y.~Wang, A.~Bayen, and Y.~Wu.
\newblock The surprising effectiveness of mappo in cooperative multi-agent games.
\newblock In \emph{Proceedings of the Neural Information Processing Systems Track on Datasets and Benchmarks}, 2022.

\end{thebibliography}

\end{document}